\documentclass[preprintnumbers,secnumarabic,amsmath,amssymb,nofootinbib,floatfix]{revtex4}
\usepackage{varioref,exscale,latexsym,amsmath,amssymb}
\usepackage{graphicx}

\usepackage{graphicx}
\usepackage{slashed}
\usepackage{dcolumn}
\usepackage{bm}

\def\beq{\begin{equation}}

\def\eeq{\end{equation}}

\def\beqa{\begin{eqnarray}}

\def\eeqa{\end{eqnarray}}

\begin{document}
\preprint{ACFI-T16-24}

\title{{\bf Is the spin connection confined or condensed? }}

\medskip\

\author{ John F. Donoghue${}$}
\email[Email: ]{donoghue@physics.umass.edu}
\affiliation{~\\
Department of Physics,
University of Massachusetts\\
Amherst, MA  01003, USA\\
 }

\begin{abstract}
The spin connection enters the theory of gravity as a nonabelian gauge field associated with local Lorentz transformations. Normally it is eliminated from making an extra assumption - that of the metricity of the vierbein field. However, treated by itself with the usual gauge action, it has a negative beta function, implying that it is asymptotically free. I suggest that the spin connection could be confined (or perhaps partially confined) in the same way as other nonabelian gauge fields. This would remove the need to make the extra assumption of metricity, as the spin connection would not be present in the low energy
theory, leaving the symmetry to be realized only using metric variables.
\end{abstract}
\maketitle

When applied to fermions, general covariance most naturally involves two fields - the vierbein (or
tetrad) $e^a_\mu (x)$ and the spin connection (or Lorentz connection) $A^{ab}_\mu(x)$. This construction,
originally due to Utiyama \cite{Utiyama:1956sy} and Kibble \cite{Kibble:1961ba}, will be reviewed below. In general this is not a purely metric theory of gravity. In order to reduce to
General Relativity, one needs to impose an extra constraint, that of {\em metricity} for the veirbein, which relates the spin connection to the usual connection given by derivatives of the metric. The relation is
\begin{equation}\label{metricity}
  A^{ab}_\mu(x) = e^{a\nu}( \partial_\mu e^b_\nu  - \Gamma_{\mu\nu}^{~~\lambda} e^b_\lambda)
\end{equation}
where $\Gamma_{\mu\nu}^\lambda $ is the usual connection defined using the metric. This eliminates the spin connection as an independent degree of freedom. This extra constraint is unfortunate because it goes beyond the simple symmetry construction based purely on general covariance.

However, the spin connection is a non-abelian gauge field of the group $SO(3,1)$. Our present expectation, based largely on lattice gauge theory, is that non-abelian gauge fields are confined. The group $SO(3,1)$ is not a compact group so that it is not totally clear that this expectation applies to the spin connection. However, its Euclidean partner is $O(4)$, which is known to be confined. If confinement does occur, it would remove the need for the metricity constraint as a separate assumption. This is because the symmetry can be realized below the confinement scale using only the vierbein, which automatically forms a metric theory. The spin connection would be removed from the low energy spectrum not by an arbitrary constraint but by confinement.

General Relativity\footnote{My conventions follow closely those of the textbook by Gasperini \cite{Gasperini}, except for $\omega^{ab}_\mu\to A^{ab}_\mu$ which I have adopted to emphasize the nature of the connection as a gauge field. The quantum field theory of gravity in this convention summarized in Ref. \cite{EPFL}.} is described by a metric field $g_{\mu\nu}(x)$. Under general coordinate changes $x^\mu \to x'^\mu$, the metric transforms in such a way that the infinitesimal distance $ds^2 = g_{\mu\nu}dx^\mu dx^\nu$ is invariant, i.e. $ g'_{\mu\nu}(x')~dx'^\mu dx'^\nu = g_{\mu\nu}(x)~dx^\mu dx^\nu$. The equivalence principle then allow us to choose a coordinate system that locally redefines the metric to be flat at a given point. The change from a general coordinate system to an infinitesimally flat one defines the vierbein field $e^i_\mu(x)$ and allows the metric to be written in terms of the vierbein as
\begin{equation}
g_{\mu\nu}(x) = \eta_{ab}~ e^a_\mu(x)~ e^b_\nu(x)
\end{equation}
where $\eta_{ab}$ is the flat Minkowski metric. One also defines the inverse metric $g^{\mu\nu}$ and inverse vierbein $e_a^\mu$ with $e_a^\mu e^a_\nu(x)= \delta^\mu_\nu $ and $e_a^\mu e^b_\mu(x)= \delta^b_a $. Latin indices are raised and lowered with $\eta^{ab},~~\eta_{ab}$ and Greek ones with $g^{\mu\nu}(x),~~g_{\mu\nu}(x)$.

In addition to the general coordinate invariance, under which the vierbein transforms as
\begin{equation}
e'^a_\mu = \frac{\partial x^\nu}{\partial x'^\mu}e^a_\nu
\end{equation}
there is extra {\em local} Lorentz symmetry
\begin{equation}
e'^a(x) = \Lambda^a_{~c}(x)~e^c(x)  ~~~~{\rm with } ~~~ \eta_{ab}~\Lambda^a_{~c}(x)~\Lambda^b_{~d}(x)  = \eta_{cd}
\end{equation}
When dealing with spinless particles this extra Lorentz symmetry (denoted in this paper by latin indices) is inessential, and all physics can be written in terms of the metric.

However, when dealing with fermions the situation is different. The Dirac matrices $\gamma^a$ do not transform as coordinates and here the $a$ index is a flat Lorentz-like index. In order to write a invariant Dirac Lagrangian the gamma matrices and the derivatives must be contracted using the (inverse) vierbein
\begin{equation}
  {\cal L} = \bar{\psi} [i \gamma^a e_a^\mu(x)  \partial_\mu +.....]\psi
\end{equation}
In addition, the fermions transform under the local Lorentz symmetry
\begin{equation}\label{localLor}
\psi \to \psi'(x') = S(x) \psi (x)
\end{equation}
where in matrix notation
\begin{equation}
S(x) =  \exp \left( \frac{-i}{2} J_{ab}\alpha^{ab}(x)\right)
\end{equation}
where $\alpha^{ab}(x) $ is the parameter associated with the local Lorentz transformation $\Lambda$ and
\begin{equation}\label{angs}
  J_{ab} = \frac{\sigma_{ab}}{2}~~~~~~~~~~~{\rm with}~~~  \sigma_{ab} =\frac{i}{2}[\gamma_a, \gamma_b]  \ \ .
\end{equation}
In order that this symmetry be local, we introduce\cite{Utiyama:1956sy, Kibble:1961ba} a gauge field $A^{ab}_\mu$ and covariant derivative $D_\mu$ with
\begin{equation}\label{fermion}
 {\cal L} = \bar{\psi} [i \gamma^a e_a^\mu(x) D_\mu -m]\psi
\end{equation}
with
\begin{equation}\label{covariant}
 D_\mu = \partial_\mu -ig \frac{J_{ab}}{2}A^{ab}_\mu  \equiv \partial_\mu -i g \mathbf{A}^{ab}_\mu
\end{equation}
Here $g$ is a coupling constant (not to be confused with the determinant of the metric), which is introduced in order to allow the field $A^{ab}_\mu$ to have the proper kinetic energy Lagrangian, which will be shown below. Traditionally the factor of g is absorbed into the field
as no kinetic term is considered.

Under the local Lorentz transformation of Eq. \ref{localLor}, the fields transform as
\begin{eqnarray}
  \mathbf{A}_\mu ' &=& S \mathbf{A}_\mu S^{-1} -\frac{i}{g}(\partial_\mu S)S^{-1}  \nonumber \\
  e_a^{\mu '} &=& \Lambda_a^{~b} (x) e_b^\mu ~~~~~ ~{\rm with }   ~~~~S^{-1}(x)\gamma^a S(x)\Lambda_a^{~b}(x)= \gamma^b
\end{eqnarray}
This combined with the general coordinate transformation
\begin{eqnarray}
  \mathbf{A'}_\mu  &=& \frac{\partial x^\nu}{\partial x'^\mu}\mathbf{A}_\nu   \nonumber \\
  e'^\mu_a &=& \frac{\partial x'^\mu}{\partial x'^\nu}e_a^\nu
\end{eqnarray}
define the symmetries of the theory.

It is straightforward to define a field strength tensor for the gauge field. We note that the spin algebra is that of $SO(3,1)$, with
the generators satisfying
\begin{equation}\label{commutator}
  [J_{ab},J_{cd}] = i \left( \eta_{ad}J_{bc}+\eta_{bc} J_{ad} -\eta_{ac} J_{bd}- \eta_{bd}J_{ac}\right)  \ \ .
\end{equation}
Equivalently, using the notation for antisymmetric indices
\begin{equation}\label{anti}
  [ab] = \frac12 (ab-ba) ~~~~~~     {\rm and } ~~~~~  a].....[b  = \frac12 (~ a.....b ~  -~ b.....a~ )
\end{equation}
we can write this as
\begin{equation}\label{commutator2}
  [J_{ab},J_{cd}] = 2i f_{[ab][cd][ef]}J^{ef}
\end{equation}
using the structure constants $f_{[ab][cd][ef]}$ defined via
\begin{eqnarray}\label{structure}
  f_{[ab][cd][ef]} &=& -\frac14 \left[ \eta_{bc}\eta_{de}\eta_{fa} - \eta_{bd}\eta_{ce}\eta_{fa} -\eta_{bc}\eta_{df}\eta_{ea} +\eta_{bd}\eta_{cf}\eta_{ea} \right. \nonumber \\
   & &\left. ~~~~~~ -\eta_{bca}\eta_{de}\eta_{fb} + \eta_{ad}\eta_{ce}\eta_{fb} +\eta_{ac}\eta_{df}\eta_{eb} -\eta_{ad}\eta_{cf}\eta_{eb} \right]  \nonumber \\
  &\equiv&  2 \eta_{b][c}\eta_{d][e}\eta_{f][a}
\end{eqnarray}
These structure constants are totally antisymmetric in the pairs $[ab],~[cd],~[ef]$. The field strength tensor is
defined via
\begin{equation}\label{commutatorD}
  [D_\mu,D_\nu] = -ig \frac{J_{ab}}{2} R^{ab}_{\mu \nu}
\end{equation}
with
\begin{equation}\label{fieldstrength}
  R^{[ab]}_{\mu\nu} = \partial_\mu A^{[ab]}_\nu - \partial_\nu A^{[ab]}_\mu +g f^{[ab]}_{~~~~[cd][ef]}  A^{[cd]}_\mu A^{[ef]}_\nu
\end{equation}
or more simply
\begin{equation}\label{fieldstrength2}
  R^{ab}_{\mu\nu} = \partial_\mu A^{ab}_\nu - \partial_\nu A^{ab}_\mu +g  ( A^{ac}_\mu ~A_{\nu c}^{~~~b} -  A^{ac}_\nu~ A_{\mu c}^{~~~b})
\end{equation}
This construction is more general than the one reviewed here involving fermions. The covariant derivative can be defined for any spin by generalizing the
spin generator $J_{ab}$, always reproducing the same field strength tensor $R^{[ab]}_{\mu\nu}$.

At this stage, there are two separate fields, the spin connection $A^{ab}_\mu(x)$ and the vierbein $e_a^\mu (x)$. Conventionally, these are
tied together by an extra assumption, one which is distinct from the symmetries of the theory. This is the {\em metric condition for the vierbein} or simply {\em metricity}.
By defining covariant derivatives in the usual way, one postulates that the covariant derivative of the vierbein vanishes. Specifically this
implies that
\begin{equation}
 \nabla_\mu e^a_\nu = 0 = \partial_\mu e^a_\nu + g A^a_{~b\mu} e^b_\nu - \Gamma_{\mu\nu}^{~~\lambda} e^a_\lambda \ \ .
\end{equation}
Here $\Gamma_{\mu\nu}^{~~\lambda} $ is the usual connection defined from the metric
\begin{equation}
\Gamma_{\mu\nu}^{~~\lambda}  = \frac12 g^{\lambda\sigma}\left[\partial_\mu g_{\sigma\nu}+ \partial_\nu g_{\mu\sigma}- \partial_\sigma g_{\mu\nu}\right]
\end{equation}
Solving for the spin connection leads to the conclusion shown
in Eq. 1 - that the spin connection is determined from the vierbein and is not
an independent field if this assumption is made\footnote{The condition can also be imposed using a first order formalism \cite{Kibble:1961ba}, but
this is simply another mechanism for making the same assumption as there are many other possible actions besides the one assumed in the first order formalism.}. Note that in Eq. 1, I reverted to the tradition of absorbing the coupling constant into the field $gA^{ab}_\mu \to A^{ab}_\mu$. After imposing metricity, and converting Lorentz indices to spacetime ones
\begin{equation}
R_{\mu\nu\alpha\beta} = e_{a\alpha}e_{b\beta} R^{ab}_{\mu\nu}
\end{equation}
we recover usual general relativity with $R_{\mu\nu\alpha\beta}$ being the Riemann tensor.

Of course, one is not forced to introduce the spin connection as an independent field. If one wants the theory to be a purely metric theory from the start, one is able to construct the Dirac Lagrangian directly with
the vierbein field, yielding the same result as occurs after the imposition of metricity.
This would be the path of the classical theory where the geometric picture
is paramount. However, given what we have learned about the construction of fundamental theories as gauge theories, the
spin connection is the most natural part of the construction. To gauge theorists, it is more clearly a fundamental field. In this case,
the imposition of the metricity condition feels unnatural.

However, if the spin connection is confined (or partially confined) and does not appear in the low energy spectrum, this extra assumption is not required. Even if the spin connection cannot propagate at low energy, the symmetry of the theory is unchanged. That symmetry can be realized using only the metric and
veirbein. In this case, the symmetry plus the lack of a propagating spin connection requires that the construction must necessarily be that of the usual metric theory.

Will the spin connection be confined? The analogy to other nonabelian gauge fields certainly suggest that confinement is possible. However, we do not have analytic control over the non-perturbative region, so that there is not a simple analytic
test to answer this question. Nevertheless indications do point to this outcome.

The spin connection treated as a gauge field is asymptotically free and describes a theory which is weakly coupled at high energies and strongly coupled at low
energies. Consider first the usual gauge Lagrangian\footnote{This is the equation which defines the normalization of the kinetic energy term. If we had absorbed the coupling constant into the spin connection, the coupling would reappear here as the initial coefficient $1/4$ would become $1/4g^2.$}
\begin{equation}\label{gauge}
{\cal L} =- \frac14 R^{ab}_{\mu\nu} R_{ab}^{\mu\nu}  \ \ .
\end{equation}
Following from this one obtains the usual Feynman rules for gauge theories, but with the gauge structure constants $f_{ijk}$ replaced by
$ f_{[ab][cd][ef]}$. All of the quantization and loop calculation goes through as usual, with
the only change being the quadratic Casimir being modified from
\begin{equation}\label{gaugecasimir}
  f_{imn}f_{jmn} = C_2 \delta_{ij}
\end{equation}
with $C_2=N$ for $SU(N)$ to
\begin{equation}\label{gravcasimir}
  f_{[ab][cd][ef]}f^{[gh][cd][ef]} = C_2 \delta^{[gh]}_{[ab]}
\end{equation}
with the result that $C_2 =2$. The resulting beta function is then
\begin{equation}
 \beta(g) ~=~ -\frac{11 C_2}{3} \frac{g^3}{16\pi^2}~=~ -\frac{22} {3} \frac{g^3}{16\pi^2}
\end{equation}
In contrast with usual gauge theories, the fermion loop does not contribute to the renormalization of the charge, as defined by the action Eq. \ref{gauge}. This will be discussed more below. Asymptotic freedom implies that the charge is weak at high energy, but grows non-perturbatively large at low energy. The apparent divergence of the coupling constant at low energy is generally treated as an indication of confinement (infrared slavery).

 An
oversimplified version of the logic for confinement is that isolated charged fields must have large fields around them by Gauss' law, costing an energy
which grows as the coupling grows.
A singlet combination of fields does not have to pay this price. Heuristically, this can be argued by noting that the interaction between
two gluons or two spin connections is attractive in the singlet channel at first order in perturbation theory. The gluons or spin connections
are massless in this approximation, and naively one can take their momentum towards zero. The resulting attractive interaction however is growing
at low energy such that this state falls below zero energy and forms a singlet condensation. Of course, this perturbative picture does not
do justice to the strong dynamics at low energy. However, our naive ways developed to conceptualize confinement and vacuum condensation in
QCD seem to apply in the singlet channel to the spin connection as well, so that it appears reasonable to consider this possibility.

It is possible that partial confinement would also have the desired effect. If the resulting vacuum condensate at strong coupling does not absolutely confine but still exhibits a large energy gap, of order the Planck energy, then spin connection excitations would not propagate at low energy and we would not be able to distinguish total confinement from partial confinement.

However, one important difference from usual gauge theories is that the group $SO(3,1)$ is non-compact. If one ignores interactions, one can see that different
components of the free field Hamiltionian enter with different signs, which can be traced back to the contraction with the flat metric tensor $\eta_{ab}$ which has terms of both signs. However, if the theory is confined or even simply strongly interacting, the free field Hamiltonaian may not have any relevance to the physical spectrum \cite{Donoghue:2017fvm}.  Since this feature is a property of the free theory, it is not clear that the interacting theory
is ill-defined. In particular, if the connection is confined, it does not itself appear in the spectrum. If one looks at the Lorentzian path-integral treatment
\begin{equation}\label{PI}
  Z= \int [d A] e^{i\int d^4x ~\left[-\frac14 R^{ab}_{\mu\nu} R_{ab}^{\mu\nu} \right] }
\end{equation}
the issue of the signs $+i$ or $-i$ in the exponent do not by themselves influence the convergence of the path integral. In order to make the integral fully well defined, one defines the Lorentzian path integral from the analytic continuation of the the Euclidean path integral. In the latter situation, one must redefine the field variables also.  The Lorentz gauge group $O(3,1)$ becomes $O(4)$ in Euclidean space. The Euclidean action carries a definite sign and the path integral becomes well defined.

If we address the issue of confinement using lattice gauge theory, the spin connection is clearly confined close to flat space. This is because the lattice theory is defined using the Euclidean continuation. This compact $O(4)$ gauge group shares the same beta function as the Lorentzian version\footnote{Note that there can be different normalization conventions for $O(N)$ groups. Ours is defined by the the Euclidean version of the covariant derivative of Eq. \ref{covariant}}. It confines in the same way as other non-Abelian groups. So if the Euclidean analytic continuation is meaningful, the spin connection will be confined.

There is a second complication which is also fundamentally important. The existence of the vierbein allows one to connect the Lorentz indices $a,~b$ and the
the spacetime indices $\mu,~\nu$. This can lead to other invariant Lagrangians beyond the obvious gauge Lagrangian of Eq. \ref{gauge}. For example, we can contract such a pair as $R^a_\mu = e_b^\nu R^{ab}_{\mu\nu}$ and add a term to the Lagrangian such as $R^a_\mu R^\mu_a$. Indeed the action linear in the curvatures could have two scalar invariants $M_1^2 e_a^\mu e_b^\nu R^{ab}_{\mu\nu}$ as well as the usual scalar curvature
$M_2^2 R (g)$ formed out of the metric. When metricity is imposed, these two objects are identical, but without that condition they are distinct. At high energies there are far more invariants.
These possible actions are not completely optional, unless limited by a symmetry. In particular, in the sense of any effective field theory  \cite{Donoghue:1994dn} they are needed for the renormalization of the theory.

As an example, a vacuum polarization diagram involving a fermion loop, coupled as in Eq. \ref{fermion}, leads to the divergences represented by
\begin{equation}
\Delta {\cal L} = - \frac{1}{384\pi^2\epsilon} \partial_a \tilde{w}_b \partial_{a'} \tilde{w}_{b'} \left[\eta^{aa'}\eta^{bb'}  -\eta^{ab'}\eta^{ba'}\right]
\end{equation}
Here we have defined a combination of the fundamental fields
\begin{equation}
\tilde{w}_d =  \epsilon_{abcd}e^{a\mu}A_\mu^{bc}
\end{equation}
A covariant field strength tensor can be constructed from this using the covariant objects
\begin{eqnarray}
E^c_{\mu\nu} &=& \nabla_\mu e^c_\nu -\nabla_\nu e^c_\nu = \partial_\mu e^c_\nu -\partial_\nu e^c_\mu + A^c_{\mu d}e^d_\nu  -  A^c_{\nu d}e^d_\mu  \nonumber \\
\tilde{N}_\mu &=& \frac12 \epsilon_{abcd} e^{a\lambda}e^{b\nu}E^c_{\lambda\nu}e^d_\mu
\end{eqnarray}
such that $\tilde{N}_\mu = ...+e^d_\mu \tilde{w}_d$. With the latter we form the covariant field strength
\begin{equation}
\tilde{N}_{\mu\nu} = \partial_\mu \tilde{N}_\nu -\partial_\nu \tilde{N}_\mu
\end{equation}
in which case the fermion loop appears as part of the covariant action
\begin{equation}\label{dual}
\Delta {\cal L} = - \frac{1}{192\pi^2\epsilon}  \frac14 \tilde{N}_{\mu\nu}\tilde{N}^{\mu\nu}
\end{equation}
This particular Lagrangian vanishes if we impose the metricity condition, and hence it is distinct from the gauge lagrangian of Eq. \ref{gauge}. However it becomes clear that the the overall parameter space to be explored is larger than in most gauge theories, as the number of possible terms in the Lagrangian is large. The possibilities of confinement or condensation may depend on the linear combinations of invariants that enter the action. More discussion of the basis of invariants can be found in \cite{Donoghue:2016xnh}.

Analytic methods are not able to address this question with rigor. At present, lattice techniques are the only available option to resolve this
question definitively. Lattice simulations of quantum gravity are notoriously complicated because one attempts to retain the full diffeomorphism invariance of the theory\cite{Loll:1998aj}. However, a tentative exploratory pathway can be identified.

For our problem a simpler question can be posed. Let us consider the gauge Lagrangian of Eq. \ref{gauge} in flat space, with the metric set equal to $\eta_{\mu\nu}$. In this construction, the confined spin connection can be readily simulated in flat space. This will remain valid for small gravitational fields, which could be added as a background field perturbation.

As a second step, one can add the vierbein field. The structure of this action was addressed in a classic paper by Tomboulis \cite{Tomboulis}. However, he added metricity as a delta function constraint constraint. This constraint should be easy to remove.

Holdom and Ren \cite{Holdom} have also recently proposed that confinement may play a role in the theory of gravity. However in their case, they are exploring
a purely metric theory with $ R +R^2 $ interactions in contrast to the discussion of the present paper which focuses on the spin connection. Their interesting suggestion also deserves to be explored more. Earlier suggestions of confinement in gravity by Smilga \cite{Smilga} also are in the context of the metric. Given the nonlinear nature of gravity and the analogies with non-abelian gauge theories, it is perhaps surprising that this possibility has been so lightly explored.

This paper discusses the possibility that the spin connection can be treated as an independent gauge field in the generally covariant theory of gravity, instead of being eliminated by the extra assumption of the metricity constraint for the vierbein. Treated independently with the usual gauge interaction, it is asymptotically free and therefore strongly coupled at low energy. This raises the natural idea that it could be confined, leaving only the metric variables at low energy. This idea can be explored without studying the full generally coordinate invariant theory, perhaps using lattice methods.

\section*{Acknowledgements} I would like to thank Eugene Golowich, Leandro Bevilaqua, Ted Jacobson, Michael Endres, Renate Loll, Pierre Ramond, Guido Martinelli, Alberto Salvio and Martin Luscher for useful conversations about this topic. This work has been supported in part by the National Science Foundation under grants NSF PHY15-20292 and NSF PHY12-25915.

\end{document}